\documentclass[12pt]{iopart}
\usepackage{iopams}
\usepackage{bm}
\usepackage{cancel}
\begin{document}
\title{One more time on the helicity decomposition of spin and orbital optical currents}
\author{Andrea Aiello}
\address{Max Planck Institute for the Science of Light, Staudtstrasse 2, 91058
Erlangen, Germany}
\ead{andrea.aiello@mpl.mpg.de}

\begin{abstract}
The helicity representation of the linear momentum density of a light wave is well understood for monochromatic optical fields in both paraxial and non-paraxial regimes of propagation.
In this note we generalize such representation to  nonmonochromatic optical  fields.  We find that, differently from the monochromatic case, the linear momentum density, aka the Poynting vector divided by $c^2$, does not separate into the sum of right-handed and left-handed terms, even when the so-called electric-magnetic democracy in enforced by averaging the electric and magnetic contributions.
However, for  quasimonochromatic light, such a separation is approximately restored after time-averaging.
This paper is dedicated to Sir Michael Berry on the occasion of his $80$th birthday.
\end{abstract}
%
%
\vspace{2pc}
\noindent{\it Keywords\/}: Helicity, non-monochromatic light, Poynting vector, spin and orbital angular momentum of light
%

\submitto{\jpa}
%
%
%

\section{Introduction}\label{intro}

Exciting ideas of Michael Berry have often inspired my work over the last two decades. On one occasion I had the privilege of working with him on a little problem concerning the helicity decomposition of spin and orbital optical currents of monochromatic light. It therefore seems appropriate to present here the natural extension of that work to the more general case of non monochromatic light waves.

Light is a complicated physical phenomenon. As Berry had shown in 2009, even the very familiar concept of energy flow, or current, becomes problematic when it comes to optical fields \cite{Berry_414}.
In this work we choose the familiar Poynting vector, which is proportional to the linear momentum density of the electromagnetic field \cite{MandelBook},  to represent the optical current. In $2007$ Bekshaev and Soskin described the decomposition of the Poynting vector of paraxial optical fields, into orbital and spin currents \cite{BEKSHAEV2007332}. Shortly afterwards, Berry could generalize such decomposition to nonparaxial monochromatic fields, by enforcing the so-called electric-magnetic democracy  \cite{Berry_414}. In the same work, elaborating on an old argument by Wolf  \cite{Wolf_1959}, Berry also showed that  the monochromatic Poynting vector separates into positive and negative helicity contributions. Subsequently, other authors have explored the subject in great detail  \cite{PhysRevA.80.063814,PhysRevA.82.063825,Bekshaev_2011,Berry_477}, so that today we can safely say that the problem has been largely understood for monochromatic light.

However, in this rather clear landscape a small landmark is still missing, namely  the helicity decomposition of spin and orbital optical currents for \emph{nonmonochromatic} optical fields. The purpose of this paper is to remedy this minor shortcoming by providing a perfectly general theory for the separation of the linear momentum density of   nonmonochromatic light waves,  into orbital and spin parts, and into positive and negative helicities. We find that for nonmonochromatic light waves the latter separation cannot be  achieved and cross-helicity terms are always present, although they become negligible for narrow-linewidth light. To obtain our results,  we use the  well-known Helmholtz decomposition theorem \cite{Panofsky}, to obtain the gauge-invariant spin and orbital parts of optical linear momentum in  real space. Then, we switch to reciprocal (Fourier) space to illustrate and clarify the role of electric-magnetic democracy for the cancellation of the cross-helicity terms occurring in the monochromatic case.

\section{Maxwell's equations in real and reciprocal spaces}\label{MaxEq}

In this section we briefly review some textbooks results about Maxwell's equations and helicity (circular polarization) basis, mainly to set the working scenario and to fix an appropriate notation \cite{CohenTann,Zangwill}.
Maxwell's equations in empty space are \cite{Jackson},
\numparts
\begin{eqnarray}
 \bm{\nabla} \cdot \mathbf{E}(\mathbf{r},t) & =  0, \label{eq10a} \\[6pt]
 \bm{\nabla} \cdot \mathbf{B}(\mathbf{r},t) & =  0, \label{eq10b} \\[4pt]
 \bm{\nabla} \times \mathbf{E}(\mathbf{r},t)  & =  -\frac{\partial}{\partial t }\mathbf{B}(\mathbf{r},t) , \label{eq10c} \\[2pt]
 \bm{\nabla} \times \mathbf{B}(\mathbf{r},t) & =  \frac{1}{c^2} \frac{\partial}{\partial t }\mathbf{E}(\mathbf{r},t)  , \label{eq10d}
\end{eqnarray}
\endnumparts
where $\mathbf{E}(\mathbf{r},t)$ and $\mathbf{B}(\mathbf{r},t)$ are the electric and magnetic fields, respectively, and $c$ is the speed of light in vacuum.
Throughout this paper,  we will denote real-valued three-vectors with upright bold characters like $\mathbf{E}$ or $\mathbf{r}$, while we will use bold Italic characters like $\bi{A}$ or $\bepsilon$, to signify complex-valued three-vectors.  Unit vectors will be distinguished from ordinary vectors by a caret, like $\hat{\mathbf{e}} \in \mathbb{R}^3$ or $\hat{\bepsilon} \in \mathbb{C}^3$, so that $\hat{\mathbf{e}} \cdot \hat{\mathbf{e}}=1$ and $\hat{\bepsilon}^* \cdot \hat{\bepsilon}=1$.

For the sake of definiteness, consider now the electric field only.
Taking the curl of  \eref{eq10c}  and using   the vector identity
\begin{eqnarray}
\bm{\nabla} \times \left( \bm{\nabla} \times \mathbf{A} \right) \, = \, \bm{\nabla} \left(\bm{\nabla} {\cdot} \mathbf{A} \right) \, - \, \nabla^{2}\mathbf{A}, \label{eq20}
\end{eqnarray}
together with \eref{eq10a}, we obtain the electromagnetic wave equation
\begin{eqnarray}
 \nabla^2  \mathbf{E}(\mathbf{r},t) - \frac{1}{c^2} \frac{\partial^2}{\partial t^2 } \, \mathbf{E}(\mathbf{r},t) =0. \label{eq30}
\end{eqnarray}
Let $\tilde{\bi{E}}(\mathbf{k},t)$ be the spatial Fourier transform of $\mathbf{E}(\mathbf{r},t)$. Then, by definition $\mathbf{E}(\mathbf{r},t)$ and $\tilde{\bi{E}}(\mathbf{k},t)$ satisfy the equations \cite{CohenTann},
\numparts
\begin{eqnarray}
\mathbf{E}(\mathbf{r},t) & =   \int \frac{\rmd^3 k}{\left( 2 \pi \right)^{3/2}} \, \tilde{\bi{E}}(\mathbf{k},t) \exp \left(  i  \mathbf{k} \cdot \mathbf{r} \right) , \label{eq40a} \\[4pt]
 \tilde{\bi{E}}(\mathbf{k},t) & =   \int \frac{\rmd^3 r}{\left( 2 \pi \right)^{3/2}} \, \mathbf{E}(\mathbf{r},t)\exp \left( - i  \mathbf{k} \cdot \mathbf{r} \right), \label{eq40b}
\end{eqnarray}
\endnumparts
where $\mathbf{E}(\mathbf{r},t) \in \mathbb{R}^3 $ implies  $ \tilde{\bi{E}}(\mathbf{-k},t) = \tilde{\bi{E}}^*(\mathbf{k},t)$, and from \eref{eq10a}  it follows that $\mathbf{k} \cdot \tilde{\bi{E}}(\mathbf{k},t)=0$. Throughout this paper, it is understood that  integrals in both real and reciprocal space are taken over the whole $\mathbb{R}^3$-space.
Substituting \eref{eq40a} into \eref{eq30}, we obtain
\begin{equation}
\left( \frac{\partial^2}{\partial t^2 } + \omega^2 \right)  \tilde{\bi{E}}(\mathbf{k},t) = 0,  \label{eq50}
\end{equation}
where here and hereafter $\omega = c |\mathbf{k}| =  c \, k$, being $k = 2 \pi/\lambda$ the wavenumber of the light of wavelength $\lambda$. The solution of the second-order differential equation \eref{eq50} is
\begin{equation}
\tilde{\bi{E}}(\mathbf{k},t) = \bi{a}(\mathbf{k}) \exp \left( - i \omega t \right) +  \bi{a}^*(-\mathbf{k}) \exp \left(  i \omega t \right),  \label{eq60}
\end{equation}
where the vector amplitude $\bi{a}(\mathbf{k})$ is determined by the value of the field and of its time derivative at $t =0$, as follows:
\begin{eqnarray}
\bi{a}(\mathbf{k}) &  = \frac{1}{2} \left[\tilde{\bi{E}}(\mathbf{k},t) + \frac{i}{\omega} \frac{\partial}{\partial t} \tilde{\bi{E}}(\mathbf{k},t) \right]_{t = 0}\nonumber \\[6pt]
& = \frac{1}{2}  \int \frac{\rmd^3 r}{\left( 2 \pi \right)^{3/2}} \left[ \mathbf{E}(\mathbf{r},t) + \frac{i}{\omega} \frac{\partial}{\partial t} \, \mathbf{E}(\mathbf{r},t) \right]_{t=0} \exp \left( - i \mathbf{k} \cdot \mathbf{r}\right), \label{eq70}
\end{eqnarray}
where \eref{eq40b} has been used.
However, for reasons that will be soon clear, it is more convenient to rewrite $\bi{a}(\mathbf{k}) $ in terms of \emph{time-dependent} fields, as
\begin{eqnarray}
\bi{a}(\mathbf{k}) & = \frac{1}{2}   \int \frac{\rmd^3 r}{\left( 2 \pi \right)^{3/2}} \left[ \mathbf{E}(\mathbf{r},t) + \frac{i}{\omega} \frac{\partial}{\partial t} \, \mathbf{E}(\mathbf{r},t) \right] \exp \left( - i \mathbf{k} \cdot \mathbf{r} + i \omega t \right) \, . \label{eq75}
\end{eqnarray}
 It is not difficult to show (see \ref{app1}),  that $\bi{a}(\mathbf{k})$ defined by \eref{eq75} is actually time-independent, so that it coincides with the definition \eref{eq70}.
Then, substituting \eref{eq60} into \eref{eq40a}, we obtain
\begin{eqnarray}
\mathbf{E}(\mathbf{r},t)  =  \mathrm{Re} \left\{ 2 \int \frac{\rmd^3 k}{\left( 2 \pi \right)^{3/2}} \, \bi{a}(\mathbf{k}) \exp \left[  i  \left( \mathbf{k} \cdot \mathbf{r} - \omega t \right) \right] \right\}. \label{eq80}
\end{eqnarray}

All the results above following \eref{eq20} apply, \textit{mutatis mutandis}, to the magnetic field $\mathbf{B}(\mathbf{r},t)$ and its spatial Fourier transform  $\tilde{\bi{B}}(\mathbf{k},t)$.
Therefore, we can write
\begin{eqnarray}
\mathbf{B}(\mathbf{r},t) & =   \int \frac{\rmd^3 k}{\left( 2 \pi \right)^{3/2}} \, \tilde{\bi{B}}(\mathbf{k},t) \exp \left(  i  \mathbf{k} \cdot \mathbf{r} \right), \label{eq90}
\end{eqnarray}
where
\begin{equation}
\tilde{\bi{B}}(\mathbf{k},t) = \frac{1}{c} \Bigl[ \bi{b}(\mathbf{k}) \exp \left( - i \omega t \right) +  \bi{b}^*(-\mathbf{k}) \exp \left(  i \omega t \right) \Bigr],  \label{eq95}
\end{equation}
with
\begin{eqnarray}
\bi{b}(\mathbf{k}) & = \frac{c}{2} \int \frac{\rmd^3 r}{\left( 2 \pi \right)^{3/2}} \left[ \mathbf{B}(\mathbf{r},t) + \frac{i}{\omega} \frac{\partial}{\partial t} \, \mathbf{B}(\mathbf{r},t) \right] \exp \left( - i \mathbf{k} \cdot \mathbf{r} + i \omega t \right) \, . \label{eq105}
\end{eqnarray}
 Finally, substituting \eref{eq95} into \eref{eq90}, we obtain
\begin{eqnarray}
\mathbf{B}(\mathbf{r},t)  =  \mathrm{Re} \left\{ \frac{2}{c} \int \frac{\rmd^3 k}{\left( 2 \pi \right)^{3/2}} \, \bi{b}(\mathbf{k}) \exp \left[  i  \left( \mathbf{k} \cdot \mathbf{r} - \omega t \right) \right] \right\}. \label{eq107}
\end{eqnarray}

The reciprocal-space vector amplitudes $\bi{a}(\mathbf{k})$ and $\bi{b}(\mathbf{k})$ of the electric and magnetic fields, respectively,  are not independent because of \eref{eq10c} and \eref{eq10d}. Substituting \eref{eq80} and \eref{eq107} into \eref{eq10c} and \eref{eq10d} we find the two equivalent equations
\numparts
\begin{eqnarray}
\hat{\mathbf{k}} \times \bi{a}(\mathbf{k}) -  \bi{b}(\mathbf{k})  & =  0 , \label{eq120a} \\[4pt]
 \bi{a}(\mathbf{k}) +  \hat{\mathbf{k}} \times \bi{b}(\mathbf{k}) & = 0, \label{eq120b}
\end{eqnarray}
\endnumparts
where $\hat{\mathbf{k}} = \mathbf{k}/ |\mathbf{k}|$.  From \eref{eq10a} and \eref{eq10b} it follows that
\begin{eqnarray}
\mathbf{k} \cdot \bi{a}(\mathbf{k}) = \, 0 \, = \mathbf{k} \cdot \bi{b}(\mathbf{k}) \, , \label{eq110}
\end{eqnarray}
and \eref{eq120a} and \eref{eq120b} imply
\begin{eqnarray}
\bigl|\bi{a}(\mathbf{k})\bigr|^2 = \bigl|\bi{b}(\mathbf{k})\bigr|^2 \, . \label{eq125}
\end{eqnarray}
Using (\ref{eq120a}-\ref{eq120b}) and \eref{eq125}, we can calculate the total energy $U$ of the electromagnetic field:
\begin{eqnarray}
U  & =  \frac{\varepsilon_0}{2} \int \rmd^3 r  \left[ \mathbf{E}(\mathbf{r},t) \cdot \mathbf{E}(\mathbf{r},t) + c^2 \mathbf{B}(\mathbf{r},t) \cdot \mathbf{B}(\mathbf{r},t) \right]  , \nonumber \\[4pt]
 & = \varepsilon_0 \int \rmd^3 k  \, \Bigl\{ \left|\bi{a}(\mathbf{k})\right|^2 + \left|\bi{b}(\mathbf{k})\right|^2 \Bigr\} . \label{eq127}
\end{eqnarray}

It is often convenient rewriting \eref{eq120a} and \eref{eq120b} in a Cartesian coordinate system with the three perpendicular axes parallel to the unit vectors $\hat{\mathbf{e}}_1(\hat{\mathbf{k}}), \hat{\mathbf{e}}_2(\hat{\mathbf{k}})$ and $\hat{\mathbf{e}}_3(\hat{\mathbf{k}}) = \hat{\mathbf{k}}$, such that
\begin{eqnarray}
\hat{\mathbf{e}}_a(\hat{\mathbf{k}}) \cdot \hat{\mathbf{e}}_b(\hat{\mathbf{k}}) = \delta_{ab}, \qquad \mathrm{and} \qquad
\hat{\mathbf{e}}_a(\hat{\mathbf{k}}) \times \hat{\mathbf{e}}_b(\hat{\mathbf{k}}) = \varepsilon_{abc} \, \hat{\mathbf{e}}_c(\hat{\mathbf{k}}) , \label{eq130}
\end{eqnarray}
where $\varepsilon_{abc}$ denotes the Levi-Civita symbol with $a,b,c \in \{ 1,2,3 \}$, and summation over repeated indices (Einstein's summation convention), is understood. Without loss of generality, we may choose
\begin{eqnarray}
\hat{\mathbf{e}}_2(\hat{\mathbf{k}})  = \frac{\hat{\mathbf{n}} \times \hat{\mathbf{k}}}{\bigl| \hat{\mathbf{n}} \times \hat{\mathbf{k}} \bigr|}, \qquad \mathrm{and} \qquad
\hat{\mathbf{e}}_1(\hat{\mathbf{k}})  = \hat{\mathbf{e}}_2(\hat{\mathbf{k}}) \times \hat{\mathbf{e}}_3(\hat{\mathbf{k}}) , \label{eq130bis}
\end{eqnarray}
where $\hat{\mathbf{n}}$ is an arbitrary real unit vector in reciprocal space. From this choice it follows that
\begin{eqnarray}
\hat{\mathbf{e}}_1(-\hat{\mathbf{k}})  = \hat{\mathbf{e}}_1(\hat{\mathbf{k}}), \quad
\hat{\mathbf{e}}_2(-\hat{\mathbf{k}})  = - \, \hat{\mathbf{e}}_2(\hat{\mathbf{k}}), \quad \mathrm{and} \quad
\hat{\mathbf{e}}_3(-\hat{\mathbf{k}})  = - \, \hat{\mathbf{e}}_3(\hat{\mathbf{k}}). \label{eq135}
\end{eqnarray}
Then, from \eref{eq110} it follows that we can always write
\numparts
\begin{eqnarray}
\bi{a}(\mathbf{k})  & =  a_1(\mathbf{k}) \, \hat{\mathbf{e}}_1(\hat{\mathbf{k}}) + a_2(\mathbf{k})  \, \hat{\mathbf{e}}_2(\hat{\mathbf{k}}), \label{eq140a} \\[4pt]
\bi{b}(\mathbf{k}) & =  b_1(\mathbf{k})  \, \hat{\mathbf{e}}_1(\hat{\mathbf{k}}) + b_2(\mathbf{k})  \, \hat{\mathbf{e}}_2(\hat{\mathbf{k}}), \label{eq140b}
\end{eqnarray}
\endnumparts
where $a_c(\mathbf{k}) = \hat{\mathbf{e}}_c(\hat{\mathbf{k}}) \cdot \bi{a}(\mathbf{k})$ and $b_c(\mathbf{k}) = \hat{\mathbf{e}}_c(\hat{\mathbf{k}}) \cdot \bi{b}(\mathbf{k})$, with $c=1,2$.
Substituting \eref{eq140a} and \eref{eq140b} into \eref{eq120a} and \eref{eq120b}, and using \eref{eq130} we obtain, after a straightforward manipulation,
\numparts
\begin{eqnarray}
a_1(\mathbf{k}) -  b_2(\mathbf{k}) & =  0 , \label{eq150a} \\[4pt]
a_2(\mathbf{k}) +  b_1(\mathbf{k}) & = 0. \label{eq150b}
\end{eqnarray}
\endnumparts
These two equations can  be suggestively rewritten in matrix notation as
\begin{eqnarray}
\left(
  \begin{array}{c}
               a_1 \\
               a_2 \\
  \end{array}
\right)
 + M
\left(
  \begin{array}{c}
               b_1 \\
               b_2 \\
  \end{array}
\right) = 0
 , \label{eq151}
\end{eqnarray}
where
\begin{eqnarray}
M =  \left(
  \begin{array}{cc}
  0 & -1 \\
  1 & 0 \\
  \end{array}
\right). \label{eq152}
\end{eqnarray}
This suggests that by diagonalizing $M$ it is possible to find  a new vector basis, say $\{ \hat{\bm{\epsilon}}_+(\hat{\mathbf{k}}), \hat{\bm{\epsilon}}_-(\hat{\mathbf{k}}) \}$,
where
\numparts
\begin{eqnarray}
\bi{a}(\mathbf{k})  & =  a_+(\mathbf{k}) \, \hat{\bm{\epsilon}}_+(\hat{\mathbf{k}}) + a_-(\mathbf{k})  \, \hat{\bm{\epsilon}}_-(\hat{\mathbf{k}}), \label{eq142a} \\[4pt]
\bi{b}(\mathbf{k}) & =  b_+(\mathbf{k})  \, \hat{\bm{\epsilon}}_+(\hat{\mathbf{k}}) + b_-(\mathbf{k})  \, \hat{\bm{\epsilon}}_-(\hat{\mathbf{k}}), \label{eq142b}
\end{eqnarray}
\endnumparts
and such that the new components
\numparts
\begin{eqnarray}
a_\sigma(\mathbf{k})  & = \hat{\bm{\epsilon}}_\sigma^*(\hat{\mathbf{k}}) \cdot \bi{a}(\mathbf{k}), \label{eq143a} \\[4pt]
b_\sigma(\mathbf{k}) & =  \hat{\bm{\epsilon}}_\sigma^*(\hat{\mathbf{k}}) \cdot \bi{b}(\mathbf{k}), \label{eq143b}
\end{eqnarray}
\endnumparts
with $\sigma = \pm 1$, are uncoupled and satisfy the equation
\begin{eqnarray}
a_\sigma(\mathbf{k}) +  \lambda_\sigma \, b_\sigma(\mathbf{k})  =  0, \qquad (\sigma = \pm 1). \label{eq144}
\end{eqnarray}
Here,  $\lambda_\sigma = i \, \sigma$ are the eigenvalues of $M$ associated with the eigenvectors
\begin{eqnarray}
\hat{\bm{\epsilon}}_\sigma(\hat{\mathbf{k}})= \frac{\hat{\mathbf{e}}_1(\hat{\mathbf{k}}) - i  \, \sigma \, \hat{\mathbf{e}}_2(\hat{\mathbf{k}})}{\sqrt{2}}, \qquad (\sigma = \pm 1). \label{eq156}
\end{eqnarray}
These two orthogonal unit complex vectors form the so-called helicity (or, circular) polarization basis \cite{Berry_414}, and have the following properties:
\numparts
\begin{eqnarray}
\hat{\mathbf{k}} \cdot \hat{\bm{\epsilon}}_{\sigma}(\hat{\mathbf{k}}) & =  0 , \label{eq280a} \\[6pt]
 \hat{\mathbf{k}} \times \hat{\bm{\epsilon}}_{\sigma}(\hat{\mathbf{k}}) & =  i \, \sigma \, \hat{\bm{\epsilon}}_{\sigma}(\hat{\mathbf{k}}), \label{eq280b} \\[6pt]
\hat{\bm{\epsilon}}_\sigma^*(\hat{\mathbf{k}}) \cdot \hat{\bm{\epsilon}}_{\sigma'}(\hat{\mathbf{k}}) &  = \delta_{\sigma \sigma'} , \label{eq280c} \\[6pt]
 \hat{\bm{\epsilon}}_{\sigma}^*(-\hat{\mathbf{k}}) & =  \hat{\bm{\epsilon}}_{\sigma}(\hat{\mathbf{k}}), \label{eq280d}
\end{eqnarray}
\endnumparts
where $\sigma, \sigma' = \pm 1$, and (\ref{eq130}-\ref{eq135}) and \eref{eq156} have been used.
It should be noticed that in the definition of the circular polarization basis \eref{eq156}, we use an opposite-sign  notation with respect to Eq. (3.22) in \cite{Berry_414}. The connection between the two notations is simply: $\hat{\bm{\epsilon}}_+(\hat{\mathbf{k}}) = \bm{e}_-$ and $\hat{\bm{\epsilon}}_-(\hat{\mathbf{k}}) = \bm{e}_+$.

\section{Helicity decomposition of the fields}\label{Hdec}

In this section, we extend the well-known helicity decomposition for monochromatic optical fields in empty space \cite{Berry_414}, to perfectly general fields. It is worth noting that the more general concept of optical helicity and its connections with the angular momentum of light and with the symmetry properties of electromagnetic fields,  have  been studied extensively in recent years \cite{Barnett_2010,PhysRevA.83.021803,PhysRevA.85.063810,Cameron_2012,PhysRevA.86.042103,Barnett_2016,PhysRevA.105.023524}.

\subsection{Local and nonlocal helicity decomposition}\label{HdecI}

To begin with, let us rewrite \eref{eq80} with respect to the helicity basis $\{ \hat{\bm{\epsilon}}_+(\mathbf{k}), \hat{\bm{\epsilon}}_-(\mathbf{k}) \}$, as
\begin{eqnarray}
\mathbf{E}(\mathbf{r},t) & =   \sum_{\sigma = \pm 1} \mathrm{Re}\left\{ 2 \int \frac{\rmd^3 k }{\left( 2 \pi \right)^{3/2}} \, \, \hat{\bm{\epsilon}}_\sigma (\hat{\mathbf{k}}) \, a_\sigma(\mathbf{k}) \exp \left[  i  \left( \mathbf{k} \cdot \mathbf{r} - \omega t \right) \right]  \right\} \nonumber \\[8pt]
 & \equiv   \mathbf{E}_+(\mathbf{r},t) + \mathbf{E}_-(\mathbf{r},t),
\label{f10}
\end{eqnarray}
where \eref{eq142a} has been used. Using \eref{eq40a} and \eref{eq40b} we can also rewrite
\begin{eqnarray}
\mathbf{E}_\sigma(\mathbf{r},t) & =   \int \frac{\rmd^3 k}{\left( 2 \pi \right)^{3/2}} \, \tilde{\bi{E}}_\sigma(\mathbf{k},t) \exp \left(  i  \mathbf{k} \cdot \mathbf{r} \right) , \label{f12}
\end{eqnarray}
where
\begin{eqnarray}
\tilde{\bi{E}}_\sigma(\mathbf{k},t) = \bi{a}_\sigma(\mathbf{k}) \exp(- i \omega t) + \bi{a}_\sigma^*(-\mathbf{k}) \exp( i \omega t), \label{f120}
\end{eqnarray}
with
\begin{eqnarray}
\bi{a}_\sigma(\mathbf{k}) = a_\sigma(\mathbf{k}) \, \hat{\bm{\epsilon}}_\sigma(\hat{\mathbf{k}})= \frac{1}{2} \left[ \bi{a}(\mathbf{k}) - i \, \sigma \, \bi{b}(\mathbf{k}) \right], \label{f140}
\end{eqnarray}
(no summation over repeated $\sigma$).

Equations (\ref{f10}-\ref{f140}) show that the helicity decomposition is \emph{local} in reciprocal space. Therefore, we expect that the corresponding decomposition in real space will have a \emph{nonlocal} (convolution) form,  as emphasized by the Bialynicki-Birulas \cite{PhysRevA.79.032112,Bialynicki_Birula_2014}. To demonstrate this, we use \eref{eq75} and \eref{eq143a}   to rewrite  $\mathbf{E}_\sigma(\mathbf{r},t)$ as
\begin{eqnarray}
\mathbf{E}_\sigma(\mathbf{r},t)   = & \mathrm{Re} \left\{
\vphantom{\frac{ \widetilde{\Theta}_\sigma(\hat{\mathbf{k}})}{c  \left| \mathbf{k} \right|}}
\int \rmd^3 r' \! \left[  \int \frac{\rmd^3 k}{\left( 2 \pi \right)^{3}}\, \widetilde{\Theta}_\sigma(\hat{\mathbf{k}}) \exp \left[  i  \mathbf{k} \cdot\left( \mathbf{r} - \mathbf{r}' \right) \right] \right]   \mathbf{E}(\mathbf{r}',t) \right. \nonumber \\[6pt]
 & \left. + \, i \! \int \rmd^3 r'   \left[  \int \frac{\rmd^3 k}{\left( 2 \pi \right)^{3}} \, \frac{ \widetilde{\Theta}_\sigma(\hat{\mathbf{k}})}{c  \left| \mathbf{k} \right|} \exp \left[  i  \mathbf{k} \cdot\left( \mathbf{r} - \mathbf{r}' \right) \right] \right]   \frac{\partial}{\partial t }  \mathbf{E}(\mathbf{r}',t) \right\},
\label{f50}
\end{eqnarray}
where we have defined the dyad ($3 \times 3 $ matrix) $\widetilde{\Theta}_\sigma(\hat{\mathbf{k}})$, as
\begin{eqnarray}
\widetilde{\Theta}_\sigma(\hat{\mathbf{k}}) \equiv  \hat{\bm{\epsilon}}_\sigma (\hat{\mathbf{k}}) \hat{\bm{\epsilon}}_\sigma^* (\hat{\mathbf{k}}), \qquad (\sigma = \pm 1).
\label{f60}
\end{eqnarray}
Next, we notice that from \eref{eq280d} it follows that
\begin{eqnarray}
\widetilde{\Theta}_\sigma^*(-\hat{\mathbf{k}})  = \widetilde{\Theta}_\sigma(\hat{\mathbf{k}}) .
\label{f70}
\end{eqnarray}
This implies that both integrals with respect to $\rmd^3 k $ in \eref{f50}, are real-valued so that the second term is identically zero and we can write
\begin{eqnarray}
\mathbf{E}_\sigma(\mathbf{r},t)   =   \int \rmd^3 r' \, \Theta_\sigma(\mathbf{r} - \mathbf{r}')   \mathbf{E}(\mathbf{r}',t) ,
\label{f80}
\end{eqnarray}
where we have defined
\begin{eqnarray}
\Theta_\sigma(\mathbf{r} - \mathbf{r}')  \equiv &  \int \frac{\rmd^3 k }{\left( 2 \pi \right)^{3}} \, \widetilde{\Theta}_\sigma(\hat{\mathbf{k}}) \exp \left[  i  \mathbf{k} \cdot\left( \mathbf{r} - \mathbf{r}' \right) \right] .
\label{f90}
\end{eqnarray}

Equation \eref{f80} gives the sought (nonlocal) helicity decomposition directly in real space. We remark that $\Theta_\sigma(\mathbf{r} - \mathbf{r}')$ is not a causal propagator, so that \eref{f80} simply expresses a convolution operation.

\subsection{Formal developments}\label{HdecII}

From the definition \eref{f60} it follows that
\begin{eqnarray}
\widetilde{\Theta}_\sigma \equiv \hat{\bm{\epsilon}}_\sigma  \hat{\bm{\epsilon}}_\sigma^*  = \frac{1}{2} \left(\hat{\mathbf{e}}_1 \hat{\mathbf{e}}_1 + \hat{\mathbf{e}}_2 \hat{\mathbf{e}}_2 \right) + \frac{i \sigma}{2} \left(\hat{\mathbf{e}}_1 \hat{\mathbf{e}}_2 - \hat{\mathbf{e}}_2 \hat{\mathbf{e}}_1 \right), \label{e210}
\end{eqnarray}
where the explicit dependence on $\hat{\mathbf{k}}$ has been omitted for clarity.
The completeness of the basis $\{ \hat{\mathbf{e}}_1, \hat{\mathbf{e}}_2,\hat{\mathbf{e}}_3 \}$ implies $\left(\hat{\mathbf{e}}_1 \hat{\mathbf{e}}_1 + \hat{\mathbf{e}}_2 \hat{\mathbf{e}}_2 \right)_{ab} = \delta_{ab} - k_a k_b/k^2 $, with $a,b =1,2$. The second quantity $\Theta \equiv \hat{\mathbf{e}}_1 \hat{\mathbf{e}}_2 - \hat{\mathbf{e}}_2 \hat{\mathbf{e}}_1$  is an antisymmetric second order tensor, which is a function of the components of $\hat{\mathbf{k}}$. The only antisymmetric second order tensor of this kind that one can build with  the components of $\hat{\mathbf{k}}$ has necessarily the form $[\Theta]_{ab} = f(k) \varepsilon_{abc} k_c/k$,
where $f(k)$ is an arbitrary function of $k = |\mathbf{k}|$, which is uniquely fixed by the requirements
\begin{eqnarray}
\Theta \, \hat{\mathbf{e}}_1 = -\hat{\mathbf{e}}_2, \qquad \Theta \, \hat{\mathbf{e}}_2 = \hat{\mathbf{e}}_1, \qquad \Rightarrow \qquad f(k) = 1 \, . \label{e220}
\end{eqnarray}
Therefore, we can eventually write
\begin{eqnarray}
\bigl[\widetilde{\Theta}_\sigma \bigr]_{ab} = \frac{1}{2} \left( \delta_{ab} - \frac{k_a k_b }{k^2} \right) + \frac{i \sigma}{2} \varepsilon_{abc} \frac{k_c}{k} \; , \qquad (a,b,c \in \{ 1,2,3 \}). \label{e230}
\end{eqnarray}
This  result  can also be found via a direct calculation using the  model \eref{eq130bis} for the triad $\{ \hat{\mathbf{e}}_1, \hat{\mathbf{e}}_2,\hat{\mathbf{e}}_3 \}$. Now, substituting \eref{e230} into  \eref{f90} we obtain
\begin{eqnarray}
\bigl[\Theta_\sigma (\mathbf{r} - \mathbf{r}')\bigr]_{ab}
 & =    \frac{1}{2 } \, \delta_{\mathrm{T} ab}\left(\mathbf{r} - \mathbf{r}' \right) \nonumber \\[6pt]
 & \quad + \frac{i \sigma}{2 }  \int  \frac{\rmd^3 k}{(2 \pi)^{3}} \, \left(  \varepsilon_{abc} \frac{k_c}{k} \right) \exp \left[ i \mathbf{k} \cdot \left(\mathbf{r} - \mathbf{r}' \right)\right],\label{e250}
\end{eqnarray}
where $ \delta_{\mathrm{T} ab}\left(\mathbf{r} - \mathbf{r}' \right)$ denotes the transverse delta-function \cite{LoudonBook}. The second term in \eref{e250} can be calculated  as
\begin{eqnarray}
\fl \qquad \int  \frac{\rmd^3 k}{(2 \pi)^{3}}  \left(  \varepsilon_{abc} \frac{k_c}{k}\right) \exp \left[ i \mathbf{k} \cdot \left(\mathbf{r} - \mathbf{r}' \right)\right] =
\frac{1}{i } \varepsilon_{abc} \frac{\partial }{\partial x_c} \int  \frac{\rmd^3 k}{(2 \pi)^{3}} \, \frac{\exp \left[ i \mathbf{k} \cdot \left(\mathbf{r} - \mathbf{r}' \right)\right]}{k} \,. \label{e260}
\end{eqnarray}
Recalling that
\begin{eqnarray}
\int  \rmd^3 k\, \frac{\exp \left[ i \mathbf{k} \cdot \left(\mathbf{r} - \mathbf{r}' \right)\right]}{k} =
 \frac{4 \pi}{\left|\mathbf{r} - \mathbf{r}' \right|^2 } \, ,\label{e270}
\end{eqnarray}
and that the transverse delta-function acts on the transverse field $\mathbf{E}(\mathbf{r}',t)$ as an ordinary delta-function, we can rewrite \eref{f80} as,
\begin{eqnarray}
\mathbf{E}_\sigma (\mathbf{r},t)  = &   \frac{1}{2}\,  \mathbf{E}(\mathbf{r},t) - \frac{\sigma}{(2 \pi)^2 } \bnabla \times \int \rmd^3 r' \, \frac{\mathbf{E}(\mathbf{r}',t)}{\left|\mathbf{r} - \mathbf{r}'  \right|^2} \nonumber \\[6pt]
 & = \frac{1}{2} \, \mathbf{E}(\mathbf{r},t)  -  \frac{2 \sigma}{(2 \pi)^2 }  \int \rmd^3 r' \, \frac{\mathbf{r}' \times \mathbf{E}(\mathbf{r} - \mathbf{r}' ,t)}{\left|\mathbf{r}' \right|^4} \, . \label{a90}
\end{eqnarray}
This equation displays the non-local character of the helicity decomposition for non-monochromatic light. It is worth noticing that \eref{a90}  can be also written as $\mathbf{E}_\pm(\mathbf{r},t) = P_\pm \mathbf{E}(\mathbf{r},t)$, where $P_\pm$ are the projector operators defined by Eq. (8) in \cite{PhysRevA.79.032112}.
The expression in the second line of \eref{a90}  shows that, notwithstanding the singularity at $|\mathbf{r}'| =0$,  the integral with respect to $\rmd^3 r'= {r'}^2 \rmd r'  \sin \theta' \rmd \theta' \rmd \phi' $ is actually finite with the proviso of performing integration over the angular variables $(\theta', \phi')$ before integration over the radial one $r'$, where $\mathbf{r'} = r' \hat{\mathbf{r}}' = r'(\sin \theta' \cos \phi', \sin \theta' \sin \phi', \cos \theta')$ in spherical coordinates.    This follows from the Taylor expansion of $ \mathbf{E}(\mathbf{r} - \mathbf{r}' ,t)$ around $|\mathbf{r}'| =0 $,
\begin{eqnarray}
\frac{\mathbf{r}' \times \mathbf{E}(\mathbf{r} - \mathbf{r}',t)}{\left|\mathbf{r}' \right|^4}  \approx &   \frac{1}{\left|\mathbf{r}' \right|^3} \, \hat{\mathbf{r}}' \times \mathbf{E}(\mathbf{r},t ) + O \left( \left|\mathbf{r}' \right|^{-2} \right) , \label{a95}
\end{eqnarray}
and from
\begin{eqnarray}
\int \rmd^3 r' \, \frac{1}{\left|\mathbf{r}' \right|^3} \, \hat{\mathbf{r}}' \times \mathbf{E}(\mathbf{r},t )  = &   \int\limits_0^{\infty}  \rmd r' \, \frac{1}{r'} \underbrace{\int\limits_0^\pi \rmd \theta'  \sin \theta'  \int\limits_0^{2 \pi} \rmd \phi'  \, \hat{\mathbf{r}}' \times \mathbf{E}(\mathbf{r},t )}_{= \, 0}.  \label{a97}
\end{eqnarray}
The integral of the remaining terms $O \left( \left|\mathbf{r}' \right|^{-2} \right)$ is automatically finite.

\subsection{Discussion}\label{discussion1}

Equation \eref{a90} is the straightforward generalization of the correspondent equation for monochromatic light given, for example, in Sec. 3 of \cite{Berry_477}. However, differently from the monochromatic case, here we have
\begin{eqnarray}
\bm{\nabla} \times  \mathbf{E}_\sigma(\mathbf{r},t) = - \, \sigma \int \frac{\rmd^3 k}{(2 \pi)^{3/2}}  \, k \, \tilde{\bi{E}}_\sigma(\mathbf{k},t)  \exp \left( i \mathbf{k} \cdot  \mathbf{r} \right).
\label{f150}
\end{eqnarray}
By simply rearranging terms, we can suggestively rewrite \eref{f150} as
\begin{eqnarray}
\bm{\nabla} \times  \mathbf{E}_\sigma(\mathbf{r},t)  =  &   - \sigma \, k_0 \, \mathbf{E}_\sigma(\mathbf{r},t) \nonumber \\[6pt]
 & - \sigma \, k_0   \int \frac{\rmd^3 k}{\left( 2 \pi \right)^{3/2}}   \left(\frac{k-k_0}{k_0} \right)  \tilde{\bi{E}}_\sigma(\mathbf{k},t)  \exp \left( i \mathbf{k} \cdot  \mathbf{r} \right) , \label{f160}
\end{eqnarray}
where $k_0>0$ is an arbitrarily chosen wavenumber.
This discordance with the monochromatic case (see, e.g., Eqs. (3.2) in  \cite{Berry_477}), is not surprising because the eigenfunctions of the curl operator are  simple  plane waves,
\begin{eqnarray}
\bm{\nabla} \times  \left[  \hat{\bm{\epsilon}}_\sigma(\mathbf{k})  \exp \left( i \mathbf{k} \cdot  \mathbf{r} \right) \right] =  - k \, \sigma    \left[  \hat{\bm{\epsilon}}_\sigma(\mathbf{k})  \exp \left( i \mathbf{k} \cdot  \mathbf{r} \right) \right], \label{f170}
\end{eqnarray}
with eigenvalues proportional to the wavenumber $k$, as extensively studied by Moses \cite{Moses}. Therefore, only superpositions of monochromatic  plane waves can yield eigenfunctions of the curl operator.  However,  when $ \bigl| \tilde{\bi{E}}_\sigma(\mathbf{k},t) \bigr|$ is sharply peaked at $k=k_0$,   the second term inside the integral  \eref{f160}  is $O \left( (k-k_0)/{k_0} \right)$ and it becomes negligible with respect to the first one, so that  \eref{f160}  reduces to the approximate expression
\begin{eqnarray}
\bm{\nabla} \times  \mathbf{E}_\sigma(\mathbf{r},t) \approx   - \sigma \, k_0 \, \mathbf{E}_\sigma(\mathbf{r},t) , \label{f180}
\end{eqnarray}
which becomes exact for monochromatic light of frequency $\omega_0 = c k_0$.

Finally, we notice that the connection between the helicity decomposition and the Riemann-Silberstein vectors \cite{PhysRevA.67.062114,Kaiser_2004,Bialynicki_Birula_2013} survives  in the non-monochromatic case, as already suggested by \eref{f140} in reciprocal space. In fact, substituting \eref{eq75} and \eref{eq105}  into \eref{f140}, we obtain
\begin{eqnarray}
\bi{a}_\sigma(\mathbf{k})  =  \frac{1}{2}  \int \frac{\rmd^3 r}{\left( 2 \pi \right)^{3/2}}  \left[ \bi{F}_\sigma (\mathbf{r},t) + \frac{i}{\omega} \, \frac{\partial}{\partial t} \,\bi{F}_\sigma (\mathbf{r},t)  \right] \exp\left[-i \left( \mathbf{k} \cdot  \mathbf{r} - \omega t \right) \right] , \label{f190}
\end{eqnarray}
where we have defined
\begin{eqnarray}
 \bi{F}_\sigma (\mathbf{r},t) = \frac{1}{2} \Bigl[ \mathbf{E}(\mathbf{r},t) - i \, \sigma \, c \, \mathbf{B}(\mathbf{r},t) \Bigr] . \label{f200}
\end{eqnarray}

The results presented in this section, although interesting, are hardly surprising because we simply performed some linear operations upon the fields, so that  there cannot be a profound difference between the monochromatic and the general cases. The only significant change is illustrated by \eref{f160}.
 However, when considering physical quantities represented by bilinear forms of the fields, like the linear momentum, we expect significative differences between the two cases, especially because time-averaging is required to build observable quantities for monochromatic light. This will be the subject of the next section.

\section{Helicity representation of the linear momentum}\label{LinMom}

\subsection{Spin and orbital linear momentum density in real space}\label{LinMom2}

The total linear momentum $\mathbf{P}$ of the electromagnetic field is given by
\begin{eqnarray}
\mathbf{P} =  \varepsilon_0 \int \rmd^3 r \, \mathbf{E}(\mathbf{r},t) \times \mathbf{B}(\mathbf{r},t)   , \label{p10}
\end{eqnarray}
and is connected to the Poynting vector $\mathbf{S}$ by the simple relation $\mathbf{S}  = c^2 \mathbf{P}$ \cite{MandelBook}. The linear momentum density $\mathbf{p}(\mathbf{r},t)$ is defined by
\begin{eqnarray}
\mathbf{p}(\mathbf{r},t) =  \varepsilon_0 \, \mathbf{E}(\mathbf{r},t) \times \mathbf{B}(\mathbf{r},t)   . \label{p20}
\end{eqnarray}
Substituting \eref{eq80} and \eref{eq107} into \eref{p10} we obtain,
\begin{eqnarray}
\mathbf{P} & =   \frac{2 \, \varepsilon_0}{c}  \int \rmd^3 k \, \hat{\mathbf{k}} \left| \bi{a}(\mathbf{k})\right|^2, \label{p30}
\end{eqnarray}
where \eref{eq120a} has been used, and
\begin{eqnarray}
\left| \bi{a}(\mathbf{k})\right|^2
=  \left| a_1(\mathbf{k})\right|^2 + \left| a_2(\mathbf{k})\right|^2
=  \left| a_+(\mathbf{k})\right|^2 + \left| a_-(\mathbf{k})\right|^2 . \label{p40}
\end{eqnarray}
This equation shows that the total linear momentum $\mathbf{P}$ trivially separates into the sum of the two polarization components, irrespective of the polarization basis one chooses, namely $\mathbf{P} = \mathbf{P}_1 + \mathbf{P}_2 = \mathbf{P}_+ + \mathbf{P}_-$. However, for the linear momentum density $\mathbf{p}(\mathbf{r},t)$ things are not so simple, because such separation occurs only in the circular polarization basis and when the light is paraxial or monochromatic   \cite{BEKSHAEV2007332,Berry_414}. An open question is whether this  feature will also apply to the  linear momentum density of non-paraxial and non-monochromatic light or not.

To answer this question, we begin by writing  $\mathbf{p}(\mathbf{r},t)$ in the equivalent but more symmetric form
\begin{eqnarray}
\mathbf{p}(\mathbf{r},t) = \frac{\varepsilon_0}{2}  \left[ \mathbf{E}(\mathbf{r},t) \times \mathbf{B}(\mathbf{r},t) - \mathbf{B}(\mathbf{r},t) \times \mathbf{E}(\mathbf{r},t) \right]. \label{p50}
\end{eqnarray}
In the realm of classical physics, this is just restyling, because the two terms of the sum in \eref{p50} are actually identical. However, in quantum physics this symmetrization is necessary to obtain the linear momentum density operator in a Hermitian form \cite{MandelBook}. Assuming that the fields and their spatial  derivatives vanish at infinity, we can use the Helmholtz decomposition theorem \cite{Panofsky} and Maxwell's equations \eref{eq10c} and \eref{eq10d} to write the electric and magnetic fields as
\numparts
\begin{eqnarray}
\mathbf{E} (\mathbf{r},t) &  =  -\frac{1}{4 \pi c} \, \frac{\partial}{\partial t} \int \rmd^3 r' \, \frac{\bnabla' \times c \, \mathbf{B}(\mathbf{r}',t)}{\left|\mathbf{r} - \mathbf{r}'  \right|}, \label{p60a}  \\[6pt]
c \, \mathbf{B} (\mathbf{r},t) & =  \frac{1}{4 \pi c} \, \frac{\partial}{\partial t} \int \rmd^3 r' \, \frac{\bnabla' \times \mathbf{E}(\mathbf{r}',t)}{\left|\mathbf{r} - \mathbf{r}'  \right|},  \label{p60b}
\end{eqnarray}
\endnumparts
where here and hereafter $\bnabla'$ denotes the gradient with respect to the primed coordinates $\mathbf{r}' = (x',y',z')$. These equations are correct, as it can be easily verified substituting \eref{eq80} and \eref{eq107} into both sides of \eref{p60a} and \eref{p60b}, and using \eref{eq120a} and \eref{eq120b}. Then, substituting \eref{p60a} and \eref{p60b} into \eref{p50} we obtain
\begin{eqnarray}
\mathbf{p}(\mathbf{r},t) = &  \frac{\varepsilon_0}{2} \frac{1}{4 \pi c^2} \left\{ \int \rmd^3 r' \, \frac{1}{\left|\mathbf{r} - \mathbf{r}'  \right|} \, \mathbf{E}(\mathbf{r},t) \times \left[ \bnabla' \times \frac{\partial}{\partial t} \, \mathbf{E} (\mathbf{r}',t) \right] \right. \nonumber \\[6pt]
& \left. + \, c^2 \int \rmd^3 r' \, \frac{1}{\left|\mathbf{r} - \mathbf{r}'  \right|} \, \mathbf{B}(\mathbf{r},t) \times \left[ \bnabla' \times \frac{\partial}{\partial t} \, \mathbf{B} (\mathbf{r}',t) \right] \right\}. \label{p70}
\end{eqnarray}
Now we recall the vector identity
\begin{eqnarray}
\mathbf{A}(\mathbf{r}) \times \left[ \bnabla' \times \mathbf{B}(\mathbf{r}')  \right] = & \mathbf{A}(\mathbf{r}) \cdot \left( \bnabla ' \right) \mathbf{B}(\mathbf{r}') + \bnabla' \times \left[ \mathbf{A}(\mathbf{r}) \times \mathbf{B}(\mathbf{r}')\right] \nonumber \\[6pt]
& - \mathbf{A}(\mathbf{r}) \left[ \bnabla' \cdot \mathbf{B}(\mathbf{r}') \right] , \label{p80}
\end{eqnarray}
where here and hereafter we use the  suggestive notation \cite{Berry_414},
\begin{eqnarray}
\mathbf{A} \cdot \left( \bnabla \right) \mathbf{B} = \sum_{a=1}^3  A_a \left( \bnabla B_a \right) . \label{p110}
\end{eqnarray}
Using \eref{p80} with $\bnabla' \cdot \mathbf{B}(\mathbf{r}') = 0$, we can rewrite \eref{p70} as
\begin{eqnarray}
\mathbf{p}(\mathbf{r},t) & =    \mathbf{p}_{\mathrm{orb}} (\mathbf{r},t) + \mathbf{p}_{\mathrm{sp}} (\mathbf{r},t) \nonumber \\[6pt]
& =   \frac{1}{2} \Bigl[ \mathbf{p}_{\mathrm{orb}E} (\mathbf{r},t) + \mathbf{p}_{\mathrm{orb}B} (\mathbf{r},t) \Bigr] + \frac{1}{2} \Bigl[ \mathbf{p}_{\mathrm{sp}E} (\mathbf{r},t) + \mathbf{p}_{\mathrm{sp}B} (\mathbf{r},t) \Bigr]  , \label{p90}
\end{eqnarray}
where
\numparts
\begin{eqnarray}
\mathbf{p}_{\mathrm{orb}E} (\mathbf{r},t) &  =    \frac{\varepsilon_0}{4 \pi c^2}  \int \rmd^3 r' \, \frac{1}{\left|\mathbf{r} - \mathbf{r}'  \right|} \, \mathbf{E}(\mathbf{r},t) \cdot \left( \bnabla' \right)\frac{\partial}{\partial t} \, \mathbf{E} (\mathbf{r}',t)  \, , \label{p100a}  \\[6pt]
\mathbf{p}_{\mathrm{sp}E} (\mathbf{r},t) & =   \frac{\varepsilon_0}{4 \pi c^2}  \int \rmd^3 r' \, \frac{1}{\left|\mathbf{r} - \mathbf{r}'  \right|} \,\bnabla' \times \left[ \mathbf{E}(\mathbf{r},t) \times \frac{\partial}{\partial t} \, \mathbf{E} (\mathbf{r}',t) \right] \, .  \label{p100b}
\end{eqnarray}
\endnumparts
The magnetic densities $\mathbf{p}_{\mathrm{orb}B} (\mathbf{r},t)$ and $\mathbf{p}_{\mathrm{sp}B} (\mathbf{r},t)$ are obtained from \eref{p100a} and \eref{p100b}, respectively, replacing $\mathbf{E}$ with $c \mathbf{B}$ everywhere.
We make four remarks.
\begin{enumerate}
  \item An explicit expression in the reciprocal space for the decomposition $\mathbf{p}(\mathbf{r},t) =    \mathbf{p}_{\mathrm{orb}E} (\mathbf{r},t) + \mathbf{p}_{\mathrm{sp}E} (\mathbf{r},t)$ has been given by Chun-Fang Li \cite{PhysRevA.80.063814,PhysRevA.93.049902}, following Darwin's analogous derivation for the angular momentum of light, which is based on the Fourier transform of the electromagnetic field \cite{Darwin_1932}. Our derivation in real space has the advantage to make clear the similarity with the monochromatic case. Moreover, our results are manifestly  gauge-invariant, because the  separation of the linear momentum density into its orbital and spin parts is achieved directly in terms of the measurable electric and magnetic fields.
  \item The product $\mathbf{E}(\mathbf{r},t) \times \frac{\partial}{\partial t} \, \mathbf{E} (\mathbf{r}',t) $  in \eref{p100b} reveals the spin nature of this term (cf. with Eq. (5) in \cite{Uehara_1988}).
  \item Equations \eref{p100a} and \eref{p100b} display clearly the nonlocal nature of the orbital-spin decomposition, as pointed out in \cite{PhysRevA.79.032112,Bialynicki_Birula_2014,Bialynicki_Birula_2011}. However, the whole issue of locality is, in my opinion, a bit tricky. I will give a simple example to illustrate how potentially complicated is the situation. In the Coulomb gauge, the electric field in vacuum in absence of free charges and free currents, can be written in terms of the transverse vector potential $\mathbf{A}(\mathbf{r},t)$ as $\mathbf{E}(\mathbf{r},t) = - \partial \mathbf{A}(\mathbf{r},t)/ \partial t$ \cite{LoudonBook}. By definition, the electric field is  a local physical observable. Since $\mathbf{B} = \bnabla \times \mathbf{A}$, we can use the Helmholtz theorem to write
\begin{eqnarray}
\mathbf{A}(\mathbf{r},t) &  =  \frac{1}{4 \pi}  \int \rmd^3 r' \, \frac{\bnabla' \times \mathbf{B}(\mathbf{r}',t)}{\left|\mathbf{r} - \mathbf{r}'  \right|}, \label{p102}
\end{eqnarray}
and this quantity is apparently nonlocal, according to Bialynicki-Birula\&Bialynicka-Birula (see the end of Sec. 4 in \cite{Bialynicki_Birula_2011}). However, tacking minus the time derivative of both sides of the (nonlocal) equation \eref{p102}, we obtain the (local) electric field, as given by \eref{p60a}. Thus, we arrived at an apparent paradox. How to solve this paradox (if there is a paradox at all), is not clear to me.
\item
It is worth noting that the pair of equations \eref{p60a} and \eref{p60b}, gives an additional reason for considering the Riemann-Silberstein vector $\bi{F}_\sigma(\mathbf{r},t) = [\mathbf{E}(\mathbf{r},t) - i \, \sigma \, c \, \mathbf{B}(\mathbf{r},t)]/2$ defined by  \eref{f200}. Multiplying \eref{p60b} by $-i \sigma$, with $\sigma = \pm 1$, and summing to \eref{p60a}, we obtain the eigenvalue equation
\begin{eqnarray}
 \frac{1}{4 \pi c} \, \frac{\partial}{\partial t} \int \rmd^3 r' \, \frac{1}{\left|\mathbf{r} - \mathbf{r}'  \right|}  \bnabla' \times  \bi{F}_\sigma (\mathbf{r}',t) &  = i \, \sigma \, \bi{F}_\sigma(\mathbf{r},t) \, , \label{p104}
\end{eqnarray}
which shows that $\bi{F}_\sigma(\mathbf{r},t)$ is an eigenvector of the integro-differential operator
\begin{eqnarray}
 \frac{1}{4 \pi c} \, \frac{\partial}{\partial t} \int \rmd^3 r' \, \frac{1 }{\left|\mathbf{r} - \mathbf{r}'  \right|} \, \bnabla' \times  \, , \label{p106}
\end{eqnarray}
associated with the eigenvalue $\lambda_\sigma = i \sigma$ defined by \eref{eq144}. This operator admits a very simple interpretation: it is the time derivative of the inverse curl operator, here denoted by $(\bnabla \times)^{-1}$. Such meaning comes from the fact that, by definition, $\bi{F}_\sigma(\mathbf{r},t)$ satisfies Maxwell's equations, namely
\begin{eqnarray}
 \frac{1}{c} \, \frac{\partial}{\partial t} \bi{F}_\sigma(\mathbf{r},t) = i \, \sigma \, \bnabla \times \bi{F}_\sigma(\mathbf{r},t)  \, . \label{p108}
\end{eqnarray}
Applying on both sides of this equation $(\bnabla \times)^{-1}$, we obtain again \eref{p104}. See also \cite{PhysRevA.78.052116} and \cite{Barnett_2014} for other interesting uses of the Riemann-Silberstein vector in optics.

\end{enumerate}

\subsection{Helicity decomposition of spin and orbital linear momentum density in reciprocal space}\label{HelDec}

The helicity decomposition of the optical currents $ \mathbf{p}_{\mathrm{orb}} (\mathbf{r},t)$ and $\mathbf{p}_{\mathrm{sp}} (\mathbf{r},t)$ is straightforwardly achieved in reciprocal space. Substituting \eref{eq80} and \eref{eq107} into \eref{p90}, we obtain after a little calculation,
\begin{eqnarray}
\! \! \! \! \!  \mathbf{p}_{\mathrm{orb}} (\mathbf{r},t)  & = &  \frac{1}{2} \Bigl[ \mathbf{p}_{\mathrm{orb}E} (\mathbf{r},t) + \mathbf{p}_{\mathrm{orb}B} (\mathbf{r},t) \Bigr] \nonumber \\[6pt]
&= & -  \frac{ \epsilon_0}{ c \, (2 \pi)^3} \sum_{\sigma,\sigma' = \pm 1} \int \rmd^3 k \int \rmd^3 k' \,  \exp\left[i \left(\mathbf{k} - \mathbf{k}' \right)\cdot \mathbf{r} \right] \hat{\mathbf{k}}' \nonumber \\[6pt]
&& \times  \bi{a}_{\sigma}(\mathbf{k}) \cdot \Biggl\{ \bi{a}_{\sigma'}(-\mathbf{k}') \left(\frac{ 1 - \sigma \sigma'}{2} \right) \exp \left[- i \left(\omega + \omega' \right) t \right] \nonumber \\[6pt]
&& \phantom{ \times  \bi{a}_{\sigma}(\mathbf{k}) \cdot \Biggl\{ } -    \bi{a}_{\sigma'}^*(\mathbf{k}') \left( \frac{1 + \sigma \sigma'}{2} \right) \exp \left[ - i \left(\omega - \omega' \right) t \right]\Biggr\} + \mathrm{c.c.} \,, \label{p120}
\end{eqnarray}
and
\begin{eqnarray}
\! \! \! \! \! \! \! \! \mathbf{p}_{\mathrm{sp}} (\mathbf{r},t)  & = & \frac{1}{2} \Bigl[ \mathbf{p}_{\mathrm{sp}E} (\mathbf{r},t) + \mathbf{p}_{\mathrm{sp}B} (\mathbf{r},t) \Bigr] \nonumber \\[6pt]
&= &     \frac{\epsilon_0}{c \, (2 \pi)^3} \sum_{\sigma,\sigma' = \pm 1} \int \rmd^3 k \int \rmd^3 k' \,  \exp\left[i \left(\mathbf{k} - \mathbf{k}' \right)\cdot \mathbf{r} \right]  \nonumber \\[6pt]
&& \times \hat{\mathbf{k}}' \cdot \bi{a}_{\sigma}(\mathbf{k}) \Biggl\{ \bi{a}_{\sigma'}(-\mathbf{k}') \left( \frac{1 - \sigma \sigma' }{2}\right) \exp \left[- i \left(\omega + \omega' \right) t \right] \nonumber \\[6pt]
&& \phantom{\times \hat{\mathbf{k}}' \cdot \bi{a}_{\sigma}(\mathbf{k}) \Biggl\{} -  \bi{a}_{\sigma'}^*(\mathbf{k}') \left(\frac{ 1 + \sigma \sigma'}{2} \right) \exp \left[- i \left(\omega - \omega' \right) t \right] \Biggr\} + \mathrm{c.c.} \,, \label{p130}
\end{eqnarray}
where \eref{f140} and \eref{eq144} have been used, $\omega' = c |\mathbf{k}'|$, and $\mathrm{c.c.}$ stands for c\emph{omplex} c\emph{onjugate}. Notice that the term between curly brackets is the same in both \eref{p120} and \eref{p130}. Moreover, we have used the Fourier transform relation
\begin{eqnarray}
\int  \rmd^3 r' \, \frac{\exp \left( i \mathbf{k}' \cdot \mathbf{r}' \right)}{\left| \mathbf{r} - \mathbf{r}' \right|} = \exp \left(  i \mathbf{k}' \cdot \mathbf{r} \right)
 \frac{4 \pi}{\left|\mathbf{k}'  \right|^2 } \, . \label{p140}
\end{eqnarray}
There are several aspects of \eref{p120} and \eref{p130} that are worth highlighting.
\begin{enumerate}
  \item
Differently from the monochromatic case the cross terms, which are proportional to $1 - \sigma \sigma'$, do not cancel because
\begin{eqnarray}
\frac{1 - \sigma \sigma'}{2} = \left\{
                       \begin{array}{ll}
                         0, & \sigma = \sigma' , \\[6pt]
                         1, & \sigma \neq \sigma' .
                       \end{array}
                     \right.
 \label{p150}
\end{eqnarray}
Vice versa, the diagonal terms are proportional to $1 + \sigma \sigma'$, so that
\begin{eqnarray}
\frac{1 + \sigma \sigma'}{2} = \left\{
                       \begin{array}{ll}
                         1, & \sigma = \sigma' , \\[6pt]
                         0, & \sigma \neq \sigma' .
                       \end{array}
                     \right.
 \label{p152}
\end{eqnarray}
In both expressions of $\mathbf{p}_{\mathrm{orb}} (\mathbf{r},t)$ and $\mathbf{p}_{\mathrm{sp}} (\mathbf{r},t)$, the cross terms always multiply $\exp \left[ \pm i \left(\omega + \omega' \right) t \right]$, while the diagonal terms always multiply $\exp \left[ \pm i \left(\omega - \omega' \right) t \right]$. This means that cross terms oscillate faster than the diagonal ones, so that we can neglect them  when the light is nearly monochromatic.
  \item
Using the results from \ref{MonochWaves}, we find that for monochromatic light of frequency $\omega_0$, the linear momentum densities  $\mathbf{p}_{\mathrm{orb}} (\mathbf{r}) \equiv \overline{\mathbf{p}_{\mathrm{orb}} (\mathbf{r},t)} $ and $\mathbf{p}_{\mathrm{sp}} (\mathbf{r}) \equiv \overline{\mathbf{p}_{\mathrm{sp}} (\mathbf{r},t)}$, averaged over one period $T_0 = 2 \pi /\omega_0$, are given by
\begin{eqnarray}
\mathbf{p}_{\mathrm{orb}} (\mathbf{r})  &= &   \frac{ \epsilon_0}{ c \, (2 \pi)^3} \sum_{\sigma = \pm 1} \int \rmd^3 k   \int \rmd^3 k' \,  \exp\left[i \left(\mathbf{k} - \mathbf{k}' \right)\cdot \mathbf{r} \right] \nonumber \\[6pt]
&& \times  \tilde{\bi{E}}_{\sigma}(\mathbf{k}) \cdot \tilde{\bi{E}}_{\sigma}^*(\mathbf{k}') ( \hat{\mathbf{k}} + \hat{\mathbf{k}}' ) \,, \label{p160}
\end{eqnarray}
and
\begin{eqnarray}
\mathbf{p}_{\mathrm{sp}} (\mathbf{r})  &= &   \frac{ \epsilon_0}{ c \, (2 \pi)^3} \sum_{\sigma = \pm 1} \int \rmd^3 k   \int \rmd^3 k' \,  \exp\left[i \left(\mathbf{k} - \mathbf{k}' \right)\cdot \mathbf{r} \right]  \nonumber \\[6pt]
&& \times  [ \tilde{\bi{E}}_{\sigma}(\mathbf{k}) \times \tilde{\bi{E}}_{\sigma}^*(\mathbf{k}') ] \times ( \hat{\mathbf{k}} - \hat{\mathbf{k}}' ) \, , \label{p162}
\end{eqnarray}
where $\tilde{\bi{E}}_{\sigma}(\mathbf{k}) = \hat{\bm{\epsilon}}_\sigma(\hat{\mathbf{k}}) \left[ \hat{\bm{\epsilon}}_\sigma^*(\hat{\mathbf{k}}) \cdot \tilde{\bi{E}}(\mathbf{k}) \right]$, with $\tilde{\bi{E}}(\mathbf{k}) = \bm{\alpha}(\mathbf{k})\delta \left( k - k_0 \right)$ given by \eref{eq207}.
Similar expressions for $\mathbf{p}_{\mathrm{orb}} (\mathbf{r}) $ and $\mathbf{p}_{\mathrm{sp}} (\mathbf{r})$ in two-dimensional (transverse) reciprocal space, were obtained in \cite{PhysRevA.82.063825,Bekshaev_2011}, but neglecting the contribution of evanescent waves. Conversely, the results \eref{p160} and \eref{p162} are  based on three-dimensional Fourier transform and are exact.
  \item
The origin of the factors $(1 - \sigma \sigma' )/2 = 1 - \delta_{\sigma \sigma'} $ and $(1 + \sigma \sigma')/2 = \delta_{\sigma \sigma'}$, key to the separation into the sum of the two helicities in the monochromatic case, is the  electric-magnetic democracy \cite{Berry_414,Berry_477}. It is not difficult to see that  the first term ``$1$'' in $(1 \pm \sigma \sigma' )/2$, comes from the electric field only, while the second term ``$ \pm \sigma \sigma'$'', is due to the magnetic field, because from \eref{eq144} it follows that,
\begin{eqnarray}
\fl b_\sigma(\mathbf{k}) b_{\sigma'}(\mathbf{k}') = -\sigma \sigma ' a_\sigma(\mathbf{k}) a_{\sigma'}(\mathbf{k}') \quad \mathrm{and} \quad b_\sigma(\mathbf{k}) b_{\sigma'}^*(\mathbf{k}') = \sigma \sigma ' a_\sigma(\mathbf{k}) a_{\sigma'}^*(\mathbf{k}')  \, . \label{p165}
\end{eqnarray}
  \item It is straightforward to show that
\begin{eqnarray}
 \int \rmd^3 r \, \mathbf{p}_{\mathrm{orb}} (\mathbf{r},t) = \frac{2 \, \varepsilon_0}{c}  \int \rmd^3 k \, \hat{\mathbf{k}} \left\{ \left| \bi{a}_+(\mathbf{k})\right|^2 + \left| \bi{a}_-(\mathbf{k})\right|^2 \right\} = \mathbf{P} \, ,  \label{p170}
\end{eqnarray}
and that
\begin{eqnarray}
 \int \rmd^3 r \, \mathbf{p}_{\mathrm{sp}} (\mathbf{r},t) = 0  \, , \label{p180}
\end{eqnarray}
as it should be.
The cancellation of the cross terms in \eref{p170} is due to the fact that spatial integration generates the three-dimensional  delta function $\delta(\mathbf{k} - \mathbf{k}' )$, such that
\begin{eqnarray}
\fl \delta(\mathbf{k} - \mathbf{k}' ) \, \bi{a}_{\sigma}(\mathbf{k}) \cdot \bi{a}_{\sigma'}(-\mathbf{k}') \left( \frac{1 - \sigma \sigma'}{2} \right)  & =  \delta(\mathbf{k} - \mathbf{k}' ) \, a_{\sigma}(\mathbf{k})  a_{\sigma'}(-\mathbf{k}) \, \delta_{\sigma \sigma'} \left( \frac{1 - \sigma \sigma'}{2} \right) \nonumber \\[6pt]
 & =0 \, , \label{p190}
\end{eqnarray}
where \eref{eq280c} and  \eref{eq280d} have been used. The spin part in \eref{p180} trivially cancels because of the transverse nature of the electromagnetic field, namely
\begin{eqnarray}
\delta(\mathbf{k} - \mathbf{k}' ) \, \hat{\mathbf{k}}' \cdot \bi{a}_{\sigma}(\mathbf{k}) =0 \, . \label{p200}
\end{eqnarray}
\end{enumerate}

\subsection{Helicity decomposition of total linear momentum density in reciprocal space}\label{HelDec2}

The last step required to complete our investigation is to see whether the linear momentum density $\mathbf{p} (\mathbf{r},t)$ calculated using either the electric or magnetic field, separates into right-handed and left-handed helicity contributions, or not.
A lengthy but straightforward calculation gives
\begin{eqnarray}
 \mathbf{p} (\mathbf{r},t)  & = &   \mathbf{p}_{\mathrm{orb}E} (\mathbf{r},t) + \mathbf{p}_{\mathrm{sp}E} (\mathbf{r},t)  \nonumber \\[6pt]
&= &   \frac{i \, \epsilon_0}{ c \, (2 \pi)^3} \sum_{\sigma,\sigma' = \pm 1} \int \rmd^3 k \int \rmd^3 k' \,  \exp\left[i \left(\mathbf{k} - \mathbf{k}' \right)\cdot \mathbf{r} \right]\bi{a}_{\sigma}(\mathbf{k})  \nonumber \\[6pt]
&& \times  \Biggl\{ \bi{a}_{\sigma'}(-\mathbf{k}') \left(\frac{  \sigma' - \sigma}{2} \right) \exp \left[- i \left(\omega + \omega' \right) t \right] \nonumber \\[6pt]
&& \phantom{ \times \Biggl\{ } -    \bi{a}_{\sigma'}^*(\mathbf{k}') \left( \frac{ \sigma' +  \sigma}{2} \right) \exp \left[ - i \left(\omega - \omega' \right) t \right]\Biggr\} + \mathrm{c.c.} \,, \label{p220}
\end{eqnarray}
where \eref{eq280b} and \eref{eq280d} have been used. A few things are worth noticing.
\begin{enumerate}
  \item The cross-helicity terms do not cancel. They are proportional to
\begin{eqnarray}
\frac{\sigma' - \sigma}{2} = \sigma' \left( 1 - \delta_{\sigma \sigma'} \right)\,, \label{p230}
\end{eqnarray}
and multiply the rapidly oscillating factors $\exp \left[ \pm i \left(\omega + \omega' \right) t \right]$, which average out only in the monochromatic limit (see previous discussion). Vice versa, the diagonal terms are proportional to
\begin{eqnarray}
\frac{\sigma' + \sigma}{2} = \sigma' \, \delta_{\sigma \sigma'} \,, \label{p240}
\end{eqnarray}
and survive in the monochromatic limit because they are slowly varying in time.
  \item Replacing $\bi{a}_{\sigma}(\mathbf{k})$ with $\bi{b}_{\sigma}(\mathbf{k})$ and the like in \eref{p220}, we obtain  $ \mathbf{p}_{\mathrm{orb}} (\mathbf{r},t)  =  \mathbf{p}_{\mathrm{orb}B} (\mathbf{r},t) + \mathbf{p}_{\mathrm{sp}B} (\mathbf{r},t)$.  Then, using \eref{p165} and noticing that
\begin{eqnarray}
\pm \, \sigma \sigma' \left( \frac{\sigma' \pm \sigma}{2} \right) = \frac{\sigma' \pm \sigma}{2} \,, \label{p250}
\end{eqnarray}
we find that $ \mathbf{p}_{\mathrm{orb}B} (\mathbf{r},t) + \mathbf{p}_{\mathrm{sp}B} (\mathbf{r},t) =  \mathbf{p}_{\mathrm{orb}E} (\mathbf{r},t) + \mathbf{p}_{\mathrm{sp}E} (\mathbf{r},t)$. This simply means that electric-magnetic democracy is spontaneously accomplished.
  \item Spatial integration of \eref{p220} gives again \eref{p30}, as it should be, so the result is trivial. However, it is worth highlighting the mechanism leading to the cancellation of the cross-helicity terms. In this case, differently from \eref{p190},  the key step is
\begin{eqnarray}
\hat{\bm{\epsilon}}_\sigma(\hat{\mathbf{k}}) \times \hat{\bm{\epsilon}}_{\sigma'}(-\hat{\mathbf{k}}) = \hat{\bm{\epsilon}}_\sigma(\hat{\mathbf{k}}) \times \hat{\bm{\epsilon}}_{\sigma'}^*(\hat{\mathbf{k}}) = i \, \hat{\mathbf{k}} \left( \frac{\sigma' + \sigma}{2} \right)\,, \label{p260}
\end{eqnarray}
so that
\begin{eqnarray}
\left( \frac{\sigma' + \sigma}{2} \right) \left( \frac{\sigma' \pm \sigma}{2} \right) = \delta_{\sigma \sigma'}\,, \label{p270}
\end{eqnarray}
and the cross terms disappear.
\end{enumerate}

\section{Concluding remarks}\label{conc}

The helicity decomposition of spin and orbital optical currents of a \emph{monochromatic} optical field in empty space, is a well-established fact since more than a decade \cite{BEKSHAEV2007332,Berry_414}. The aim of this paper has been to fill a small gap still existing in the current literature, by presenting in a coherent  fashion the separation of the linear momentum density of a  \emph{nonmonochromatic} electromagnetic field,  into orbital and spin parts, and into positive and negative helicities.
The three main results of this work can be summarized as follows.
\begin{enumerate}
  \item I have shown that  the Helmholtz theorem (\ref{p60a}-\ref{p60b}) permits to obtain the orbital-spin separation of the linear momentum density of a light beam directly in real space (\ref{p100a}-\ref{p100b}), without using the Fourier representation. This makes evident both the gauge invariance and the similarity with the monochromatic case.
  \item I have clarified the connection between electric-magnetic democracy and rapidly- and slowly-oscillating terms in the  helicity decomposition of $\mathbf{p}_{\mathrm{orb}}(\mathbf{r},t)$ and $\mathbf{p}_{\mathrm{sp}}(\mathbf{r},t)$, see  (\ref{p120}-\ref{p130}) and (\ref{p150}-\ref{p152}). For non-monochromatic light, cross-helicity terms do not cancel but becomes negligible for narrow-linewidth light, (e.g., lasers).
  \item I have demonstrated that for the total linear momentum density $\mathbf{p}(\mathbf{r},t)$, the electric-magnetic democracy is automatically fulfilled because $\mathbf{p}(\mathbf{r},t)= \mathbf{p}_{\mathrm{orb}B} (\mathbf{r},t) + \mathbf{p}_{\mathrm{sp}B} (\mathbf{r},t) =  \mathbf{p}_{\mathrm{orb}E} (\mathbf{r},t) + \mathbf{p}_{\mathrm{sp}E} (\mathbf{r},t)$. As in point (ii),  cross-helicity terms do not cancel but becomes negligible for narrow-linewidth light after time average.
\end{enumerate}

Finally, it is worth noting that all our results about the \emph{linear} momentum density $\mathbf{p}(\mathbf{r},t)$ can be straightforwardly extended to the  total \emph{angular} momentum density $\mathbf{j}(\mathbf{r},t) = \left(\mathbf{r} - \mathbf{r}_0 \right) \times  \mathbf{p}(\mathbf{r},t)$ of the electromagnetic field with respect to the point $\mathbf{r}_0$.

\ack
I acknowledge support from the Deutsche Forschungsgemeinschaft Project No. 429529648-
TRR 306 QuCoLiMa (``Quantum Cooperativity of Light and Matter'').

\appendix

\section{Time-independent vector field amplitude in reciprocal space}\label{app1}

We show that the vector field amplitude $\bi{a}(\mathbf{k})$ defined by \eref{eq75},  namely
\begin{eqnarray}
\bi{a}(\mathbf{k})  = \frac{1}{2}   \int \frac{\rmd^3 r}{\left( 2 \pi \right)^{3/2}} \left[ \mathbf{E}(\mathbf{r},t) + \frac{i}{\omega} \frac{\partial}{\partial t} \, \mathbf{E}(\mathbf{r},t) \right] \exp \left( - i \mathbf{k} \cdot \mathbf{r} + i \omega t \right) \, , \label{a10}
\end{eqnarray}
is time-independent, that is
\begin{eqnarray}
\frac{\partial}{\partial t} \, \bi{a}(\mathbf{k})  = 0 \, . \label{a20}
\end{eqnarray}
This follows directly from \eref{eq50}, because
\begin{eqnarray}
\frac{\partial}{\partial t} \, \bi{a}(\mathbf{k})  & = & \frac{\exp \left(  i \omega t \right)}{2}   \int \frac{\rmd^3 r}{\left( 2 \pi \right)^{3/2}} \exp \left( - i \mathbf{k} \cdot \mathbf{r}  \right) \Biggl\{ \left[ \cancel{\frac{\partial}{\partial t} \mathbf{E}(\mathbf{r},t)} + \frac{i}{\omega} \frac{\partial^2}{\partial t^2} \, \mathbf{E}(\mathbf{r},t) \right] \nonumber \\[8pt]
& & + i \omega \left[ \mathbf{E}(\mathbf{r},t) + \cancel{ \frac{i}{\omega} \frac{\partial}{\partial t} \, \mathbf{E}(\mathbf{r},t)} \right] \Biggr\} \nonumber \\[8pt]
& = &  \frac{i \exp \left(  i \omega t \right)}{2 \, \omega } \Biggl\{ \frac{\partial^2}{\partial t^2} \int \frac{\rmd^3 r}{\left( 2 \pi \right)^{3/2}} \exp \left( - i \mathbf{k} \cdot \mathbf{r} \right)\mathbf{E}(\mathbf{r},t) \nonumber \\[8pt]
&  & + \omega^2 \int \frac{\rmd^3 r}{\left( 2 \pi \right)^{3/2}} \exp \left( - i \mathbf{k} \cdot \mathbf{r} \right)\mathbf{E}(\mathbf{r},t) \Biggr\}
\nonumber \\[8pt]
& = &  \frac{i \exp \left(  i \omega t \right)}{2 \, \omega }  \left[  \frac{\partial^2}{\partial t^2}\, \tilde{\bi{E}}(\mathbf{k},t)+ \omega^2 \tilde{\bi{E}}(\mathbf{k},t)  \right] \nonumber \\[8pt]
& = & 0 \, . \label{a30}
\end{eqnarray}

\section{Monochromatic light}\label{MonochWaves}

The electric field of monochromatic light with frequency $\omega_0 > 0$ and wavenumber $k_0 = \omega_0/c$, can be written as
\begin{eqnarray}
\mathbf{E}(\mathbf{r},t)  =  \bi{E}(\mathbf{r}) \exp \left( - i \omega_0 t \right)   + \bi{E}^*(\mathbf{r}) \exp \left(  i \omega_0 t \right)  \label{eq160},
\end{eqnarray}
where  $\bi{E}(\mathbf{r})$ fulfills the Helmholtz equation:
\begin{eqnarray}
\left( \nabla^2  + k_0^2 \right) \bi{E}(\mathbf{r})  = 0. \label{eq170}
\end{eqnarray}
Allowing for generalized functions, we can write $\bi{E}(\mathbf{r})$ in terms of its spatial Fourier transform $\tilde{\bi{E}}(\mathbf{k})$, as follows:
\begin{eqnarray}
\bi{E}(\mathbf{r})  =  \int \frac{\rmd^3 k}{\left( 2 \pi \right)^{3/2}}  \, \tilde{\bi{E}}(\mathbf{k}) \exp \left( i  \mathbf{k} \cdot \mathbf{r} \right). \label{eq180}
\end{eqnarray}
Substituting \eref{eq180} into \eref{eq170}, we find that $\tilde{\bi{E}}(\mathbf{k})$ satisfies the  equation
\begin{eqnarray}
\tilde{\bi{E}}(\mathbf{k}) \left( k^2  - k_0^2 \right)    = 0, \qquad (k \equiv |\mathbf{k} |) \, , \label{eq190}
\end{eqnarray}
and it may therefore be written in the form \cite{Bogoliubov},
\begin{eqnarray}
\tilde{\bi{E}}(\mathbf{k})  = \bi{E}(\mathbf{k}) \,   \delta \left( k^2  - k_0^2 \right) \, , \label{eq200}
\end{eqnarray}
where $\bi{E}(\mathbf{k})$ is a smooth vector field in  reciprocal space. Since
\begin{eqnarray}
  \delta \left( k^2  - k_0^2 \right) = \frac{1}{2 k_0 } \, \delta \left( k  - k_0 \right)  \,   , \label{eq202}
\end{eqnarray}
for $k, k_0 \geq 0$ \cite{Hanson},  we can rewrite \eref{eq200} as
\begin{eqnarray}
\tilde{\bi{E}}(\mathbf{k}) & = \frac{\bi{E}(\mathbf{k})}{2 k_0 } \, \delta \left( k  - k_0 \right) \nonumber \\[6pt]
& \equiv \bm{\alpha}(\mathbf{k}) \,  \delta \left( k  - k_0 \right)  \,  . \label{eq207}
\end{eqnarray}
Substituting \eref{eq180} with \eref{eq207} into \eref{eq160}, we obtain
\begin{eqnarray}
\mathbf{E}(\mathbf{r},t)  =    \frac{1}{\left( 2 \pi \right)^{3/2}} \int \rmd^3 k \,  \bm{\alpha}(\mathbf{k})\, \delta \left( k  - k_0 \right) \exp \left[  i  \left( \mathbf{k} \cdot \mathbf{r} - \omega t \right) \right] + \mathrm{c.c.} \, , \label{eq205}
\end{eqnarray}
where $\mathrm{c.c.}$ stands for c\emph{omplex} c\emph{onjugate}. Comparing \eref{eq80} with \eref{eq205} we find
\begin{eqnarray}
\bi{a}(\mathbf{k})=  \bm{\alpha}(\mathbf{k}) \, \delta \left( k  - k_0 \right)   \,  . \label{eq220}
\end{eqnarray}

Similarly, the magnetic field associated with the monochromatic electric field \eref{eq160}, can be written as
\begin{eqnarray}
\mathbf{B}(\mathbf{r},t)  & =  \bi{B}(\mathbf{r}) \exp \left( - i \omega_0 t \right)   + \bi{B}^*(\mathbf{r}) \exp \left(  i \omega_0 t \right)   \nonumber \\[6pt]
& =  \frac{1}{\left( 2 \pi \right)^{3/2} } \int \rmd^3 k \, \frac{1}{c} \, \bi{b}(\mathbf{k}) \exp \left[  i  \left( \mathbf{k} \cdot \mathbf{r} - \omega t \right) \right] + \mathrm{c.c.}   \,  ,
\label{eq230}
\end{eqnarray}
where
\begin{eqnarray}
\bi{b}(\mathbf{k})=  \bm{\beta}(\mathbf{k}) \, \delta \left( k  - k_0 \right)  \, , \label{eq240}
\end{eqnarray}
with
\begin{eqnarray}
 \bm{\beta}(\mathbf{k}) = \hat{\mathbf{k}} \times  \bm{\alpha}(\mathbf{k})  \,  . \label{eq242}
\end{eqnarray}
Equations \eref{eq220} and \eref{eq240} show that  monochromatic waves are physically unrealistic because they would carry an infinite amount of energy. Indeed, substituting \eref{eq220} and \eref{eq240} into \eref{eq127}, we would obtain
\begin{eqnarray}
U  =  \varepsilon_0 \,  \int \rmd^3 k  \, \left[\delta \left( k  - k_0 \right)\right]^2  \Bigl\{ \left|\bm{\alpha}(\mathbf{k})\right|^2 + \left|\bm{\beta}(\mathbf{k})\right|^2 \Bigr\}  \,  , \label{eq260}
\end{eqnarray}
which is ill-defined because of the square of the delta function. To make sense of this expression and the like, we follow a standard regularization procedure from quantum field theory (see, e.g., $\S 64$ of \cite{Landau_4}), to write
\begin{eqnarray}
\left[\delta \left( k  - k_0 \right)\right]^2 & = c \, \delta \left( k  - k_0 \right) \delta \left( \omega  - \omega_0 \right) \nonumber \\[6pt]
& \to c \, \delta \left( k  - k_0 \right) \,\int\limits_{-T/2}^{T/2} \frac{\rmd t}{2 \pi} \, \exp \left[ i \left( \omega  - \omega_0 \right) t \right] \nonumber \\[6pt]
 & = \delta \left( k  - k_0 \right)   \frac{c \, T}{2 \pi} \, , \label{eq262}
\end{eqnarray}
where $T>0$. Then, we define the total energy per unit of time (power) $P$ of the light field, as
\begin{eqnarray}
P & =  \lim_{T \to \infty}  \, \frac{U}{T}  \nonumber \\[6pt]
 & = \frac{\varepsilon_0 \, c }{2 \pi  }\,  \int \rmd^3 k  \, \delta \left( k  - k_0 \right)  \Bigl\{ \left|\bm{\alpha}(\mathbf{k})\right|^2 + \left|\bm{\beta}(\mathbf{k})\right|^2 \Bigr\}, \label{eq264}
\end{eqnarray}
which is a finite quantity even in the limit $T \to \infty$, which is understood to be taken at the end of the calculation.

\section*{References}


\providecommand{\newblock}{}

\end{document}